\documentclass[11pt]{article}

\usepackage[margin=1in]{geometry}
\usepackage{amsmath,amssymb,amsthm}
\usepackage{graphicx}
\usepackage{booktabs}
\usepackage{caption}
\usepackage{subcaption}
\usepackage{cite}
\usepackage{xcolor}
\usepackage{indentfirst}
\usepackage{multirow}
\usepackage{pgfplots}
\usepackage{pgfplotstable}
\usepackage{tikz}
\pgfplotsset{compat=1.14}
\usepackage{hyperref}

\definecolor{facered}{HTML}{FFBBBB}
\definecolor{facegreen}{HTML}{BBFFBB}
\definecolor{faceblue}{HTML}{BBDDFF}
\definecolor{edgered}{HTML}{CC0000}
\definecolor{edgegreen}{HTML}{008800}
\definecolor{edgeblue}{HTML}{0055CC}

\theoremstyle{definition}
\newtheorem{definition}{Definition}
\newtheorem{theorem}{Theorem}
\newtheorem{proposition}{Proposition}
\newtheorem{example}{Example}

\newcommand{\Stab}{\mathcal{S}}  
\newcommand{\stab}{S}            
\newcommand{\Pauli}{\mathcal{P}}
\newcommand{\ket}[1]{|#1\rangle}

\title{Erasure Thresholds for Hyperbolic and Semi-Hyperbolic Surface Codes}
\author{Aygul Azatovna Galimova\textsuperscript{1}\\[4pt]
\textsuperscript{1}\textit{Department of Mathematics}\\
\textit{Duke University, Durham, NC 27708, USA}}
\date{}

\begin{document}

\maketitle

\begin{abstract}
We extend the circuit-level erasure noise model and Wang et al.\ quadratic expansion fitting of Chang et al.\ from planar surface codes to hyperbolic CSS surface codes. Under Chang et al.'s noise models, the $\{8,3\}$ Bolza fine-grained family threshold reaches $3.6\%$ under the general-Pauli models (which coincide at temporal resolution $\eta = 1$) and $4.7\%$ under the tailored spatially perfect model at $R_e = 1$ (pure erasure). The corresponding erasure-to-Pauli ratios ($5.0\times$ and $6.5\times$) match the surface code values to within $5\%$. Per-observable crossing-point analysis at $R_e = 1$ (Models 1--3) independently yields an erasure-to-Pauli ratio of $5.3\times$. These results establish that the erasure advantage extends to hyperbolic codes.
\end{abstract}

\tableofcontents
\newpage

\section{Introduction}

Quantum error correction (QEC) protects logical quantum information by encoding it redundantly across many physical qubits. QEC performance depends on the noise model. Matching the decoder to the noise structure can improve thresholds and reduce qubit overhead. Erasure noise---in which error locations are known to the decoder---is one such structured model, and several hardware platforms now support detecting error locations before decoding.

Physical systems that convert leakage or photon loss into detectable erasure events include dual-rail photonic qubits~\cite{knill2005quantum}, alkaline-earth Rydberg atoms~\cite{wu2022erasure}, and superconducting circuits with metastable states~\cite{kubica2023erasure, chang2024surface}. When the decoder knows which qubits have been corrupted, it can restrict its search to the erased locations. The resulting thresholds are several times higher than under standard Pauli noise.

For the toric code under code-capacity noise, the erasure threshold is $50\%$ and the depolarizing threshold is ${\sim}10.9\%$: a ratio of ${\sim}4.6\times$~\cite{dennis2002topological, stace2009thresholds}. For the planar surface code under circuit-level noise, Chang et al.\ report erasure thresholds of ${\sim}4.2\%$--$6.0\%$ depending on the erasure check model (imperfect vs.\ perfect, general vs.\ tailored decoding). The corresponding Pauli threshold is ${\sim}1.0\%$; the resulting ratios are $4.2$--$6.0\times$~\cite{chang2024surface}. Across codes and noise models, erasure thresholds consistently exceed Pauli thresholds by a factor of $4$--$5\times$.

Hyperbolic surface codes encode $k = \Theta(n)$ logical qubits into $n$ physical qubits on tessellations of closed hyperbolic surfaces, with constant encoding rate and logarithmic distance $d = O(\log n)$~\cite{breuckmann2016constructions, guth2014quantum}. In the CSS formulation, Breuckmann and Terhal~\cite{breuckmann2016constructions} establish code-capacity Pauli thresholds. Breuckmann et al.~\cite{breuckmann2017semihyperbolic} report a ${\sim}1.3\%$ phenomenological threshold for $\{4,5\}$ codes and introduce semi-hyperbolic codes via lattice subdivision. In the Floquet formulation, Fahimniya et al.~\cite{fahimniya2023faulttolerant} report phenomenological thresholds of ${\sim}0.1\%$ on $\{8,3\}$ tilings, and Higgott and Breuckmann~\cite{higgott2024constructions} achieve ${\sim}1.5$--$2\%$ under circuit-level depolarizing noise (SD6).

No prior work reports circuit-level thresholds---Pauli or erasure---for $\{p,3\}$ CSS hyperbolic surface codes. We construct 14 such codes across three tessellation families ($\{8,3\}$, $\{10,3\}$, $\{12,3\}$) and 11 semi-hyperbolic (fine-grained) codes via the Wythoff kaleidoscopic construction with the LINS algorithm~\cite{gap4, conder2002trivalent}. We focus on the $\{8,3\}$ Bolza family, which forms a proper scaling family with increasing distance under fine-graining. Results for the $\{10,3\}$ and $\{12,3\}$ families are reported in Appendix~\ref{app:other_families}.

The main findings are:
\begin{enumerate}
\item The $\{8,3\}$ base codes (6 codes, $n = 48$--$648$) achieve Pauli pseudothresholds of $0.24$--$0.49\%$, increasing with code size. Under erasure noise, three codes have measurable pseudothresholds of $3.3$--$3.7\%$; the remainder exceed the tested range.
\item The Bolza fine-grained family (H16, $f = 2$--$4$) achieves Pauli pseudothresholds up to $0.63\%$ and erasure pseudothresholds up to $4.4\%$. At $R_e = 1$ (Models 1--3), the per-observable family threshold gives an erasure-to-Pauli ratio of $5.3\times$ (Table~\ref{tab:fg_family_thresholds}), comparable to the $4.2$--$5.6\times$ reported for planar surface codes~\cite{chang2024surface}.
\item Under Chang et al.'s erasure noise model, the Bolza family threshold reaches $3.6\%$ (Models 1--3) and $4.7\%$ (Model 4) at $R_e = 1$. The erasure-to-Pauli ratios ($5.0\times$ and $6.5\times$) match the surface code values to within $5\%$.
\end{enumerate}

\section{Background}

\subsection{The Stabilizer Formalism}

\indent The stabilizer formalism~\cite{gottesman1997stabilizer} provides a group-theoretic framework for quantum error correction. For $n$ qubits, the Pauli group is $\Pauli_n = \{\pm 1, \pm i\} \times \{I, X, Y, Z\}^{\otimes n}$.

\begin{definition}[Stabilizer Code]
An $[[n, k, d]]$ stabilizer code is a $2^k$-dimensional subspace of $(\mathbb{C}^2)^{\otimes n}$ determined by an abelian subgroup $\Stab \subseteq \Pauli_n$:
\begin{equation}
\mathcal{C} = \{\ket{\psi} \in (\mathbb{C}^2)^{\otimes n} : g\ket{\psi} = \ket{\psi} \text{ for all } g \in \Stab\},
\end{equation}
where $\Stab$ does not contain $-I$, $|\Stab| = 2^{n-k}$, and $\Stab = \langle g_1, \ldots, g_{n-k} \rangle$.
\end{definition}

\begin{definition}[Logical Operators]
Logical operators are elements of the normalizer modulo the stabilizer:
\begin{equation}
\text{Logicals} = N(\Stab) / \Stab,
\end{equation}
where $N(\Stab) = \{P \in \Pauli_n : PgP^{-1} \in \Stab \text{ for all } g \in \Stab\}$.
\end{definition}

\begin{definition}[Code Distance]
The distance of a stabilizer code is:
\begin{equation}
d = \min\{\text{wt}(L) : L \in N(\Stab) \setminus \Stab\},
\end{equation}
where $\text{wt}(L)$ is the number of non-identity tensor factors of the Pauli operator $L$.
\end{definition}

\begin{definition}[CSS Code]
A Calderbank-Shor-Steane (CSS) code~\cite{calderbank1996good, steane1996multiple} is a stabilizer code whose generators partition into $X$-type (products of $X$ only) and $Z$-type (products of $Z$ only). The code is specified by two binary parity-check matrices $H_X \in \mathbb{F}_2^{r_X \times n}$ and $H_Z \in \mathbb{F}_2^{r_Z \times n}$, where each row of $H_X$ defines an $X$-stabilizer generator and each row of $H_Z$ defines a $Z$-stabilizer generator. For these to commute, the matrices must satisfy
\begin{equation}
H_X H_Z^T = 0 \quad \text{over } \mathbb{F}_2.
\end{equation}
The number of encoded logical qubits is $k = n - \mathrm{rank}(H_X) - \mathrm{rank}(H_Z)$, and the code distance is $d = \min(d_X, d_Z)$, where $d_X$ is the minimum weight of a nontrivial $X$-logical operator and $d_Z$ is defined analogously.
\end{definition}

\subsection{Hyperbolic Geometry and Tessellations}
\label{sec:hyperbolic_geometry}

\subsubsection{The Poincar\'e Disk Model}

\indent The hyperbolic plane $\mathbb{H}^2$ is a two-dimensional Riemannian manifold with constant negative Gaussian curvature $K = -1$. In the Poincar\'e disk model $\mathbb{D} = \{z \in \mathbb{C} : |z| < 1\}$, the metric is:
\begin{equation}
ds^2 = \frac{4|dz|^2}{(1-|z|^2)^2}.
\end{equation}

\indent The distance between two points $z_1, z_2 \in \mathbb{D}$ is:
\begin{equation}
\text{dist}(z_1, z_2) = 2 \text{arctanh} \left( \frac{|z_1 - z_2|}{|1 - z_1 \bar{z}_2|} \right).
\end{equation}

\indent The isometries of $\mathbb{D}$ are M\"obius transformations in $\text{PSU}(1,1)$. These transformations preserve hyperbolic distances but distort Euclidean lengths.

\subsubsection{Fuchsian Groups and Quotient Surfaces}

\indent A Fuchsian group is a discrete subgroup of $\text{PSL}(2, \mathbb{R})$, the group of orientation-preserving isometries of $\mathbb{H}^2$. Each Fuchsian group $\Gamma$ has a fundamental domain $\mathcal{F} \subset \mathbb{H}^2$ such that translates of $\mathcal{F}$ tile the hyperbolic plane.

\begin{definition}[Fundamental Domain]
A fundamental domain for a Fuchsian group $\Gamma$ is a connected region $\mathcal{F} \subset \mathbb{H}^2$ such that:
\begin{enumerate}
\item $\bigcup_{\gamma \in \Gamma} \gamma(\mathcal{F}) = \mathbb{H}^2$ (the translates cover the plane),
\item $\gamma_1(\mathcal{F})^\circ \cap \gamma_2(\mathcal{F})^\circ = \emptyset$ for $\gamma_1 \neq \gamma_2$ (interiors are disjoint).
\end{enumerate}
\end{definition}

\indent When $\Gamma$ is torsion-free (no elements of finite order except identity), the quotient $\mathbb{H}^2 / \Gamma$ is a closed orientable surface of genus $g \geq 2$. The genus is determined by the hyperbolic area of the fundamental domain via the Gauss-Bonnet theorem:
\begin{equation}
\text{Area}(\mathcal{F}) = 2\pi(2g - 2) = -2\pi\chi,
\end{equation}
where $\chi = 2 - 2g$ is the Euler characteristic.

\subsubsection{Schl\"afli Symbols and Tessellations}

\begin{definition}[Schl\"afli Symbol]
A $\{p, q\}$ tessellation is a regular tiling where each face is a regular $p$-gon and each vertex has degree $q$.
\end{definition}

\begin{theorem}[Hyperbolic Condition]
A $\{p,q\}$ tessellation exists in the hyperbolic plane if and only if:
\begin{equation}
(p-2)(q-2) > 4.
\end{equation}
Equivalently: $\frac{1}{p} + \frac{1}{q} < \frac{1}{2}$.
\end{theorem}

\indent For our code families:
\begin{itemize}
\item $\{8,3\}$: $(8-2)(3-2) = 6 > 4$.
\item $\{10,3\}$: $(10-2)(3-2) = 8 > 4$.
\item $\{12,3\}$: $(12-2)(3-2) = 10 > 4$.
\end{itemize}

\indent To obtain finite codes, we find a torsion-free normal subgroup $\Gamma \subset \Delta(p,q,2)$ of the triangle group. The finite quotient $\Delta(p,q,2)/\Gamma$ defines a tiling of the closed surface $\mathbb{H}^2 / \Gamma$ with $V$ vertices, $E$ edges, and $F$ faces~\cite{higgott2024constructions}.

\subsubsection{Quotient Surfaces and Combinatorial Parameters}

\indent A closed hyperbolic surface $\Sigma_g$ of genus $g \geq 2$ is realized as a quotient $\mathbb{H}^2 / \Gamma$, where $\Gamma$ is a discrete, torsion-free Fuchsian group. The Euler characteristic relates the cell counts:
\begin{equation}
\chi = V - E + F = 2 - 2g.
\end{equation}

\begin{proposition}[Combinatorial Parameters]
For a uniform $\{p,q\}$ tessellation of a genus-$g$ surface (all faces are $p$-gons, all vertices have degree $q$), the constraints $qV = 2E = pF$ yield:
\begin{equation}
F = \frac{4q(g-1)}{pq - 2p - 2q}, \quad E = \frac{pF}{2}, \quad V = \frac{pF}{q}.
\end{equation}
\end{proposition}

\begin{example}[Computing Parameters for $\{8,3\}$, Genus 2]
For $\{8,3\}$ on genus-2:
\begin{align}
F &= \frac{4 \cdot 3 \cdot (2-1)}{8 \cdot 3 - 2 \cdot 8 - 2 \cdot 3} = \frac{12}{24 - 16 - 6} = \frac{12}{2} = 6, \\
E &= \frac{8 \cdot 6}{2} = 24, \quad V = \frac{8 \cdot 6}{3} = 16.
\end{align}
Verification: $V - E + F = 16 - 24 + 6 = -2 = 2 - 2(2)$. \checkmark
\end{example}

\subsubsection{Asymptotic Scaling}

\indent The distinction between Euclidean and hyperbolic geometry affects quantum code parameters. In Euclidean geometry, area grows polynomially with radius ($A \sim r^2$), while in hyperbolic geometry, area grows exponentially ($A \sim e^r$).

\begin{proposition}[Euclidean Scaling]
For surface codes on Euclidean tessellations (e.g., the square lattice $\{4,4\}$):
\begin{equation}
n = \Theta(d^2), \quad k = \Theta(1), \quad \frac{k}{n} = \Theta(1/n) \to 0.
\end{equation}
The encoding rate vanishes as code size increases.
\end{proposition}

\begin{theorem}[Hyperbolic Scaling]
For hyperbolic surface codes with $n$ physical qubits~\cite{breuckmann2016constructions}:
\begin{equation}
\frac{k}{n} = \Theta(1) \quad \text{(constant rate)}, \quad d = \Theta(\log n) \quad \text{(logarithmic distance)}.
\end{equation}
For edge-based CSS codes on $\{p,q\}$ tessellations, the asymptotic rate is $1 - 2/p - 2/q$~\cite{breuckmann2016constructions}. Fine-graining (Section~\ref{sec:fine-graining}) improves the distance scaling to $d = O(\sqrt{n})$ at fixed $k$. The resulting codes are termed semi-hyperbolic.
\end{theorem}

\indent The logarithmic distance follows from systolic geometry: the systole (shortest non-contractible loop) of a hyperbolic surface of genus $g$ is at most $4\log g + O(1)$, and since $g \sim n$ for hyperbolic codes, $d = O(\log n)$~\cite{guth2014quantum}.
Figures~\ref{fig:tess_8_3}--\ref{fig:tess_12_3} illustrate three hyperbolic tessellations in the Poincar\'e disk model.

\begin{figure}[htbp]
\centering

\caption{The $\{8,3\}$ tessellation of the hyperbolic plane in the Poincar\'e disk model. Each face is a regular octagon with three faces meeting at every vertex.}
\label{fig:tess_8_3}
\end{figure}

\begin{figure}[htbp]
\centering
%
\caption{The $\{10,3\}$ tessellation. Each face is a regular decagon.}
\label{fig:tess_10_3}
\end{figure}

\begin{figure}[htbp]
\centering
%
\caption{The $\{12,3\}$ tessellation. Each face is a regular dodecagon. The exponential shrinking of faces toward the disk boundary reflects the negative curvature of the hyperbolic plane.}
\label{fig:tess_12_3}
\end{figure}

\subsection{Hyperbolic Surface Codes}

\indent For CSS surface codes on tessellations, qubits are placed on edges with stabilizers associated with faces and vertices. We write $\Stab$ for the stabilizer group and use $\stab_f$, $\stab_v$ for individual stabilizer operators (elements of $\Stab$).

\begin{definition}[Face and Vertex Stabilizers]
A CSS surface code on a $\{p,q\}$ tessellation places one physical qubit per edge, so $n = E$. For each face $f$, define
\begin{equation}
\stab_f^X = \prod_{e \in \partial f} X_e \quad (\text{weight } p).
\end{equation}
For each vertex $v$, define
\begin{equation}
\stab_v^Z = \prod_{e \in \mathrm{star}(v)} Z_e \quad (\text{weight } q).
\end{equation}
\end{definition}

\begin{proposition}[Number of Logical Qubits]
For a surface code on a genus-$g$ surface with $V$ vertices, $E$ edges, and $F$ faces:
\begin{equation}
k = E - (F-1) - (V-1) = E - F - V + 2 = 2g,
\end{equation}
using the Euler characteristic $\chi = V - E + F = 2 - 2g$. The $F$ face stabilizers satisfy one linear dependency ($\prod_f \stab_f^X = I$) and therefore have rank $F - 1$. The $V$ vertex stabilizers satisfy one dependency ($\prod_v \stab_v^Z = I$) and have rank $V - 1$~\cite{breuckmann2016constructions}.
\end{proposition}

\indent For the $\{p,3\}$ tessellations studied here, each vertex has degree $3$. The vertex stabilizers ($Z$-type) therefore have weight $3$, and the face stabilizers ($X$-type) have weight $p$ for $p \in \{8, 10, 12\}$.

\section{Methods}

\subsection{Code Construction}

\subsubsection{Wythoff Construction and Triangle Groups}

The Wythoff kaleidoscopic construction generates tessellations from reflections across the sides of a fundamental triangle~\cite{higgott2024constructions}. The triangle group $\Delta(p,q,r)$ has presentation:
\begin{equation}
\Delta(p,q,r) = \langle a, b, c \mid a^2 = b^2 = c^2 = (ab)^r = (bc)^q = (ca)^p = e \rangle,
\end{equation}
where $a, b, c$ are reflections across three sides of a triangle with interior angles $\pi/p$, $\pi/q$, $\pi/r$. For a $\{p,q\}$ tessellation, we use $\Delta(p,q,2)$, which acts on the hyperbolic plane when $1/p + 1/q + 1/2 < 1$.

A flag in the tessellation is an incident triple (vertex, edge, face). The triangle group acts simply transitively on flags, so each element of $\Delta(p,q,r)$ corresponds to exactly one flag. Vertices, edges, and faces are identified with orbits of flags under stabilizer subgroups generated by pairs of reflections.

For computation, we use the rotation subgroup $\Delta^+(p,q,2) = \langle \alpha, \beta, \gamma \mid \alpha^2 = \beta^q = \gamma^p = \alpha\beta\gamma = e \rangle$, where $\alpha = ab$, $\beta = bc$, $\gamma = ca$ are rotations around edge midpoints, vertices, and face centers respectively. In the rotation subgroup, vertices are orbits under $\langle \beta \rangle$, edges are orbits under $\langle \alpha \rangle$, and faces are orbits under $\langle \gamma \rangle$.

\subsubsection{LINS Algorithm and Face 3-Coloring}

The Low-Index Normal Subgroups (LINS) algorithm enumerates finite-index normal subgroups of finitely presented groups. We use the GAP implementation~\cite{gap4} with the LINS package to enumerate subgroups of $\Delta^+(p,3,2)$. For each normal subgroup $N$ with index $i$, the quotient $\Delta^+/N$ acts on a closed surface with $V = i/3$ vertices (since each vertex stabilizer has order 3). We filter for subgroups with genus $g \geq 2$.

The coset table produced by Todd-Coxeter enumeration encodes the graph adjacency. Geometric coordinates in the Poincar\'e disk are computed separately.

For $\{p,3\}$ tessellations constructed via the Wythoff method, the faces admit a 3-coloring from the flag structure~\cite{higgott2024constructions}. Each edge inherits its color as the complement of its two adjacent face colors:
\begin{equation}
\mathrm{color}(e) = 3 - \mathrm{color}(f_1) - \mathrm{color}(f_2),
\end{equation}
where $f_1, f_2$ are the faces incident to edge $e$ and colors are labeled $\{0,1,2\}$.

\subsubsection{Code Parameters}

We construct 14 codes across the $\{8,3\}$, $\{10,3\}$, and $\{12,3\}$ families. Table~\ref{tab:all_codes} lists the 6 $\{8,3\}$ codes; the remaining 8 codes are in Appendix~\ref{app:other_families}. Each code is named H$V$ after its vertex count $V$ (so $n = 3V/2$ physical qubits for $\{p,3\}$ tessellations). The code distance $d = \min(d_X, d_Z)$ is computed via integer linear programming (ILP), where $d_X$ is the minimum weight of a non-trivial cycle and $d_Z$ is the minimum weight of a non-trivial cocycle. Codes with $d = 2$ are omitted.

\subsection{Fine-Graining}
\label{sec:fine-graining}

\subsubsection{Procedure}

The fine-graining procedure subdivides each triangle in the dual lattice into $f^2$ smaller triangles by inserting $f-1$ vertices along each edge and $(f-1)(f-2)/2$ interior vertices per face. For a base code with parameters $[[n, k, d]]$, fine-graining with parameter $f$ produces a code with:
\begin{equation}
n' = f^2 \cdot n, \quad k' = k.
\end{equation}
The genus and number of logical qubits remain unchanged because fine-graining is a topological refinement that preserves the surface structure. The distance of the fine-grained code satisfies $d' \geq f \cdot d$; the values in Table~\ref{tab:fine_grained} are computed independently by integer linear programming. The encoding rate $k/n' = k/(f^2 n)$ decreases quadratically with $f$. Fine-graining thus trades rate for distance.

\subsubsection{Geometric Construction}

For codes derived from hyperbolic tessellations, fine-graining must respect the hyperbolic metric. New vertices along edges are placed at hyperbolic geodesic midpoints using M\"obius transformations in the Poincar\'e disk model.

Given two points $z_1, z_2 \in \mathbb{D}$, the geodesic midpoint is computed by:
\begin{enumerate}
\item Map $z_1$ to the origin via the isometry $\phi(z) = (z - z_1)/(1 - \bar{z}_1 z)$.
\item Find the hyperbolic midpoint along the radial geodesic: $w = \phi(z_2)/(1 + \sqrt{1 - |\phi(z_2)|^2})$.
\item Map back via $\phi^{-1}(w) = (w + z_1)/(1 + \bar{z}_1 w)$.
\end{enumerate}
Interior vertices are placed using hyperbolic barycentric interpolation within each triangle. This procedure ensures that the subdivided lattice inherits the correct topology from the base tessellation.

\begin{figure}[htbp]
\centering
\begin{tikzpicture}[scale=1.0]
  \node[font=\small\bfseries, anchor=south] at (-4.2,2.13) {(a) $\ell = 1$};
  \node[font=\small\bfseries, anchor=south] at (0.0,2.13) {(b) $\ell = 2$};
  \node[font=\small\bfseries, anchor=south] at (4.2,2.13) {(c) $\ell = 3$};
  \draw[->, thick, black!50] (-2.40,0.00) -- (-1.80,0.00);
  \draw[->, thick, black!50] (1.80,0.00) -- (2.40,0.00);
  \begin{scope}
    \fill[facered, fill opacity=0.3] (-5.700,-0.866) -- (-2.700,-0.866) -- (-4.200,1.732) -- cycle;
    \draw[edgeblue, semithick] (-5.700,-0.866) -- (-2.700,-0.866);
    \draw[edgered, semithick] (-2.700,-0.866) -- (-4.200,1.732);
    \draw[edgegreen, semithick] (-4.200,1.732) -- (-5.700,-0.866);
    \fill[edgeblue] (-4.200,1.732) circle(0.06);
    \fill[edgegreen] (-2.700,-0.866) circle(0.06);
    \fill[edgered] (-5.700,-0.866) circle(0.06);
  \end{scope}
  \begin{scope}
    \fill[faceblue, fill opacity=0.3] (-0.750,0.433) -- (0.750,0.433) -- (0.000,1.732) -- cycle;
    \fill[faceblue, fill opacity=0.3] (0.000,-0.866) -- (1.500,-0.866) -- (0.750,0.433) -- cycle;
    \fill[facered, fill opacity=0.3] (-1.500,-0.866) -- (0.000,-0.866) -- (-0.750,0.433) -- cycle;
    \fill[facegreen, fill opacity=0.3] (0.750,0.433) -- (-0.750,0.433) -- (0.000,-0.866) -- cycle;
    \draw[edgegreen, semithick] (-0.750,0.433) -- (0.750,0.433);
    \draw[edgeblue, semithick] (0.750,0.433) -- (0.000,1.732);
    \draw[edgered, semithick] (0.000,1.732) -- (-0.750,0.433);
    \draw[edgered, semithick] (0.000,-0.866) -- (1.500,-0.866);
    \draw[edgegreen, semithick] (1.500,-0.866) -- (0.750,0.433);
    \draw[edgeblue, semithick] (0.750,0.433) -- (0.000,-0.866);
    \draw[edgeblue, semithick] (-1.500,-0.866) -- (0.000,-0.866);
    \draw[edgered, semithick] (0.000,-0.866) -- (-0.750,0.433);
    \draw[edgegreen, semithick] (-0.750,0.433) -- (-1.500,-0.866);
    \fill[edgegreen] (0.000,1.732) circle(0.06);
    \fill[edgered] (0.750,0.433) circle(0.06);
    \fill[edgeblue] (1.500,-0.866) circle(0.06);
    \fill[edgeblue] (-0.750,0.433) circle(0.06);
    \fill[edgegreen] (0.000,-0.866) circle(0.06);
    \fill[edgered] (-1.500,-0.866) circle(0.06);
  \end{scope}
  \begin{scope}
    \fill[facegreen, fill opacity=0.3] (3.700,0.866) -- (4.700,0.866) -- (4.200,1.732) -- cycle;
    \fill[facegreen, fill opacity=0.3] (4.200,-0.000) -- (5.200,-0.000) -- (4.700,0.866) -- cycle;
    \fill[facegreen, fill opacity=0.3] (4.700,-0.866) -- (5.700,-0.866) -- (5.200,-0.000) -- cycle;
    \fill[faceblue, fill opacity=0.3] (3.200,-0.000) -- (4.200,-0.000) -- (3.700,0.866) -- cycle;
    \fill[faceblue, fill opacity=0.3] (3.700,-0.866) -- (4.700,-0.866) -- (4.200,-0.000) -- cycle;
    \fill[facered, fill opacity=0.3] (2.700,-0.866) -- (3.700,-0.866) -- (3.200,-0.000) -- cycle;
    \fill[facered, fill opacity=0.3] (4.700,0.866) -- (3.700,0.866) -- (4.200,-0.000) -- cycle;
    \fill[facered, fill opacity=0.3] (5.200,-0.000) -- (4.200,-0.000) -- (4.700,-0.866) -- cycle;
    \fill[facegreen, fill opacity=0.3] (4.200,-0.000) -- (3.200,-0.000) -- (3.700,-0.866) -- cycle;
    \draw[edgered, semithick] (3.700,0.866) -- (4.700,0.866);
    \draw[edgegreen, semithick] (4.700,0.866) -- (4.200,1.732);
    \draw[edgeblue, semithick] (4.200,1.732) -- (3.700,0.866);
    \draw[edgeblue, semithick] (4.200,-0.000) -- (5.200,-0.000);
    \draw[edgered, semithick] (5.200,-0.000) -- (4.700,0.866);
    \draw[edgegreen, semithick] (4.700,0.866) -- (4.200,-0.000);
    \draw[edgegreen, semithick] (4.700,-0.866) -- (5.700,-0.866);
    \draw[edgeblue, semithick] (5.700,-0.866) -- (5.200,-0.000);
    \draw[edgered, semithick] (5.200,-0.000) -- (4.700,-0.866);
    \draw[edgegreen, semithick] (3.200,-0.000) -- (4.200,-0.000);
    \draw[edgeblue, semithick] (4.200,-0.000) -- (3.700,0.866);
    \draw[edgered, semithick] (3.700,0.866) -- (3.200,-0.000);
    \draw[edgered, semithick] (3.700,-0.866) -- (4.700,-0.866);
    \draw[edgegreen, semithick] (4.700,-0.866) -- (4.200,-0.000);
    \draw[edgeblue, semithick] (4.200,-0.000) -- (3.700,-0.866);
    \draw[edgeblue, semithick] (2.700,-0.866) -- (3.700,-0.866);
    \draw[edgered, semithick] (3.700,-0.866) -- (3.200,-0.000);
    \draw[edgegreen, semithick] (3.200,-0.000) -- (2.700,-0.866);
    \fill[edgered] (4.200,1.732) circle(0.06);
    \fill[edgeblue] (4.700,0.866) circle(0.06);
    \fill[edgegreen] (5.200,-0.000) circle(0.06);
    \fill[edgered] (5.700,-0.866) circle(0.06);
    \fill[edgegreen] (3.700,0.866) circle(0.06);
    \fill[edgered] (4.200,-0.000) circle(0.06);
    \fill[edgeblue] (4.700,-0.866) circle(0.06);
    \fill[edgeblue] (3.200,-0.000) circle(0.06);
    \fill[edgegreen] (3.700,-0.866) circle(0.06);
    \fill[edgered] (2.700,-0.866) circle(0.06);
  \end{scope}
\end{tikzpicture}
\caption{Fine-graining a single dual triangle. Each triangle connects three face centers with distinct colors (red, green, blue). Vertex colors are assigned by barycentric interpolation modulo~3. Edge colors follow the complement rule. (a)~At $\ell = 1$, one triangle with three vertices. (b)~At $\ell = 2$, the triangle subdivides into $4$ smaller triangles with $6$ vertices. (c)~At $\ell = 3$, subdivision yields $9$ triangles with $10$ vertices. Each small triangle in the dual corresponds to a vertex in the fine-grained primal lattice.}
\label{fig:fine_graining}
\end{figure}

\subsubsection{Color Inheritance}

For the CSS codes studied here, the fine-grained lattice inherits a canonical edge coloring from the base tessellation's face 3-coloring. For subdivision parameter $f$, vertices at grid position $(i, j)$ within a triangle with corner colors $(c_A, c_B, c_C)$ receive color:
\begin{equation}
c(i,j) = c_A + i \cdot (c_C - c_A) + j \cdot (c_B - c_A) \pmod{3}.
\end{equation}
\subsection{Syndrome Extraction Circuit}

Each syndrome extraction round measures all $Z$-stabilizers and $X$-stabilizers via ancilla qubits and CX (CNOT) gates. For a $\{p,3\}$ code with $n$ data qubits, one ancilla is allocated per stabilizer generator: one per vertex ($Z$-type, weight~3) and one per face ($X$-type, weight~$p$).

The circuit for one round proceeds as follows:
\begin{enumerate}
\item Reset all ancilla qubits in $|0\rangle$.
\item Apply Hadamard to $X$-ancillas to prepare $|+\rangle$.
\item Apply CX gates between data qubits and ancillas. For $Z$-checks, data qubits are controls and the $Z$-ancilla is the target. For $X$-checks, the $X$-ancilla is the control and data qubits are targets.
\item Apply Hadamard to $X$-ancillas to return them to the $Z$ basis.
\item Measure all ancillas in the $Z$ basis.
\end{enumerate}

Two data qubits that share a stabilizer cannot interact with the same ancilla simultaneously. The CX gates are scheduled into conflict-free layers. The $Z$-checks (weight~3, on vertices) are batched into 2 groups via bipartite coloring of the vertex conflict graph. The $X$-checks (weight~$p$, on faces) are batched into 3 groups via the face 3-coloring inherited from the tessellation. Within each batch, the CX gates for each stabilizer step occupy a single layer. The total number of CX layers per round is $2 \times 3 + 3 \times p$ (6 layers for $Z$-checks, $3p$ layers for $X$-checks).

The experiment is an $X$-memory test. All data qubits are initialized in $|+\rangle$, $T = d$ syndrome extraction rounds are performed, and data qubits are measured in the $X$ basis. Detectors compare consecutive syndrome measurements. Boundary detectors compare the first (last) syndrome round to the initialization (final measurement) outcomes. The circuit tracks logical $X$ observables throughout. We simulate circuits in Stim~\cite{gidney2021stim} and decode with PyMatching~\cite{higgott2023pymatching, higgott2025sparseblossom}.

\subsection{Pauli Noise Model}

Under the Pauli noise model, two-qubit depolarizing noise of strength $p$ acts after every CX gate: each of the 15 non-identity two-qubit Paulis is applied with probability $p/15$. An $X$ error of probability $p$ acts after each ancilla reset and before each ancilla measurement. Single-qubit depolarizing noise of strength $p$ acts after each Hadamard gate. This matches the standard circuit-level depolarizing model~\cite{chang2024surface}.

\subsection{Erasure Noise Model}
\label{sec:erasure-model}

\indent An erasure error occurs when a qubit's state is lost but the loss event is detected. Erasure thresholds are higher than Pauli thresholds because the decoder can restrict its search to erased locations. We adopt the circuit-level erasure model of Chang et al.~\cite{chang2024surface}, built on the erasure qubit framework of Kubica et al.~\cite{kubica2023erasure}, parameterized by the total gate error probability~$p$ and the erasure fraction~$R_e \in [0,1]$. Each CX gate experiences one of three outcomes, mutually exclusively:
\begin{enumerate}
\item With probability $R_e \cdot p$, the gate is \emph{flagged}: both output qubits are reset to a random state (maximally mixed), and the decoder is notified (aware decoding). On the decoding graph, flagged gates contribute edges with weight zero, encoding the prior knowledge that these locations contain errors.
\item With probability $(1 - R_e) \cdot p$, an undetected Pauli error from $\{I,X,Y,Z\}^{\otimes 2} \setminus \{II\}$ is applied uniformly at random (probability $p_p / 15$ each, where $p_p = (1 - R_e) \cdot p$). These contribute standard log-likelihood-weighted edges.
\item With probability $1 - p$, the gate is noiseless.
\end{enumerate}

At $R_e = 0$ the model reduces to standard depolarizing noise. At $R_e = 1$ all errors are detected. For each shot, the specific erasure pattern is sampled and a per-shot detector error model is constructed: flagged gates contribute weight-zero edges while unflagged gates contribute edges at standard depolarizing weights. The decoder operates on this per-shot graph. We simulate at $R_e \in \{0, 0.52, 0.71, 0.91, 1.0\}$, matching Chang et al.'s experimentally motivated values.

At temporal resolution $\eta = 1$, all erasures are detected immediately after the gate, before they propagate through subsequent operations. At $\eta = 1$, three of Chang et al.'s four noise models coincide~\cite{chang2024surface}; only the tailored spatially perfect model differs, because the unleaked qubit's error type (X or~Z) is determined by the gate type (CX or CZ), providing additional information to the decoder.

For aware decoding, the per-shot approach is more expensive than batch Pauli decoding. We use 3000 shots per $(p, R_e, \text{code})$ configuration.

\subsection{Phenomenological Noise Model}
\label{sec:phenom-model}

We simulate the phenomenological noise model of Dennis et al.~\cite{dennis2002topological}: independent $X$ errors, $Z$ errors, and syndrome bit-flips each occur with probability~$p$. We decode with MWPM via PyMatching.

\subsection{Decoder}
\label{sec:decoder}

We decode $X$-type and $Z$-type errors independently using separate decoding graphs, following Chang et al.~\cite{chang2024surface}. For CSS codes with parity-check matrices $H_X$ and $H_Z$, the $X$-decoding graph is built from the $X$-stabilizer syndrome (face detectors), and the $Z$-decoding graph from the $Z$-stabilizer syndrome (vertex detectors). Each graph contains only the detectors of the corresponding type; error mechanisms that trigger detectors of both types are projected onto each graph independently.

For $\{p,3\}$ tessellations, the $X$-decoding graph is strictly a graph (no hyperedges): a single $Z$ error on a data qubit (edge) flips exactly two adjacent face stabilizers, and a measurement error on an $X$-ancilla flips two temporal detectors. No single circuit fault produces three or more $X$-type detector symptoms. The same holds for the $Z$-decoding graph by duality. This structural property ensures that minimum-weight matching and Union-Find decoders operate without information loss from hyperedge decomposition.

We use a weighted Union-Find decoder~\cite{delfosse2021almost, huang2020faulttolerant} on each decoding graph. Union-Find runs in nearly linear time and accommodates the non-uniform edge weights that arise from erasure-aware decoding: flagged (erased) edges have weight zero, while unflagged edges carry standard log-likelihood weights. The weighted variant~\cite{huang2020faulttolerant} grows clusters at rates inversely proportional to their boundary edge weights, providing an approximation to minimum-weight matching.

\subsection{Threshold Estimates}
\label{sec:threshold-methods}

\indent For each code and noise model, we sweep over a range of physical error rates and compute the per-observable logical error rate, averaged across the $k$ logical qubits. We use $T = d = \min(d_X, d_Z)$ syndrome rounds (Table~\ref{tab:all_thresholds}). We define the pseudothreshold $p^*$ as the physical error rate at which the per-cycle logical error rate $p_L = 1 - (1 - P_L)^{1/T}$ equals $k\,p$, interpolated linearly between adjacent data points; this is the break-even point where the encoded block performs as well as $k$ unencoded qubits. When $p_L$ remains below $k\,p$ throughout the tested range, we report $p^* > 5\%$. For scaling families with fixed $k$ (Table~\ref{tab:fg_family_thresholds}), we estimate family thresholds from per-observable crossing points between consecutive code sizes. These results use MWPM via PyMatching~\cite{higgott2023pymatching, higgott2025sparseblossom}.

\indent For the Chang et al.\ erasure analysis (Section~\ref{sec:threshold-vs-re}), we instead use the quadratic expansion method of Wang et al.~\cite{wang2003threshold}:
\[
P_L(p, d) \;=\; A + B\,x + C\,x^2, \qquad x = (p - p_{\mathrm{th}})\,d^{1/\nu},
\]
fitted simultaneously over all code sizes and error rates, yielding the threshold $p_{\mathrm{th}}$ and critical exponent $\nu$. These results use the weighted Union-Find decoder (Section~\ref{sec:decoder}) on the $X$-decoding graph.

\section{Results}

\subsection{Base Code Results}

\indent Table~\ref{tab:all_codes} lists the parameters of the 6 $\{8,3\}$ base codes.

\begin{table}[ht]
\centering
\caption{Parameters of the 6 $\{8,3\}$ hyperbolic surface codes. $V$ is the number of vertices (= number of $Z$-stabilizers), $n = E$ is the number of physical qubits (edges), $k = 2g$ is the number of logical qubits, and $d = \min(d_X, d_Z)$ is the code distance computed via ILP. All codes are constructed via the Wythoff kaleidoscopic construction with the LINS algorithm (Section~\ref{sec:fine-graining}). Parameters for the $\{10,3\}$ and $\{12,3\}$ families are in Appendix~\ref{app:other_families}.}
\label{tab:all_codes}
\begin{tabular}{lrrrrrr}
\toprule
Code & $V$ & $n$ & $k$ & $d$ & $k/n$ & Source \\
\midrule
H32 & 32 & 48 & 6 & 3 & 12.5\% & \cite{gap4} \\
H64 & 64 & 96 & 10 & 4 & 10.4\% & \cite{gap4} \\
H144$_{8,3}$ & 144 & 216 & 20 & 5 & 9.3\% & \cite{gap4} \\
H256 & 256 & 384 & 34 & 4 & 8.9\% & \cite{gap4} \\
H336$_{8,3}$ & 336 & 504 & 44 & 6 & 8.7\% & \cite{gap4} \\
H432$_{8,3}$ & 432 & 648 & 56 & 6 & 8.6\% & \cite{gap4} \\
\bottomrule
\end{tabular}
\end{table}

\indent Table~\ref{tab:all_thresholds} reports the Pauli and erasure pseudothresholds. The Pauli pseudothreshold increases with code size, from $0.24\%$ (H64) to $0.49\%$ (H432). Under erasure noise, three codes have measurable pseudothresholds of $3.3$--$3.7\%$; the larger codes exceed the tested range.

\begin{table}[ht]
\centering
\caption{Pauli and erasure pseudothresholds for the 6 $\{8,3\}$ hyperbolic surface codes (2000 Pauli shots, 400 erasure shots per data point). The pseudothreshold $p^*$ is defined as the solution of $p_L(p) = kp$, where $p_L = 1 - (1-P_L)^{1/T}$ is the per-cycle logical error rate. Entries marked ${>}5\%$ indicate that $p_L < kp$ throughout the tested range. Results for the $\{10,3\}$ and $\{12,3\}$ families are in Appendix~\ref{app:other_families}.}
\label{tab:all_thresholds}
\begin{tabular}{lrrrrr}
\toprule
Code & $n$ & $k$ & $T$ & $p^{*\mathrm{P}}$ & $p^{*\mathrm{E}}$ \\
\midrule
H32 & 48 & 6 & 3 & ${<}0.2\%$ & 3.33\% \\
H64 & 96 & 10 & 4 & 0.24\% & 3.66\% \\
H144$_{8,3}$ & 216 & 20 & 5 & 0.40\% & 3.63\% \\
H256 & 384 & 34 & 4 & 0.42\% & ${>}5\%$ \\
H336$_{8,3}$ & 504 & 44 & 6 & 0.47\% & ${>}5\%$ \\
H432$_{8,3}$ & 648 & 56 & 6 & 0.49\% & ${>}5\%$ \\
\bottomrule
\end{tabular}
\end{table}

\subsubsection{Threshold vs.\ Erasure Fraction}
\label{sec:threshold-vs-re}

We simulate the Bolza $\{8,3\}$ semi-hyperbolic codes (H16-f2 through H16-f5, $k=4$, $d_Z = 3,4,6,7$, $n = 64$--$400$) under the two noise models that are distinct at $\eta = 1$ (Section~\ref{sec:erasure-model}) for erasure fractions $R_e \in \{0, 0.52, 0.71, 0.91, 1.0\}$. The number of syndrome extraction rounds is $T = d_X$ for each code, where $d_X$ is the cycle distance ($d_X = 6, 8, 10, 14$ for $f = 2, 3, 4, 5$). We use 3000 shots per data point, the weighted Union-Find decoder on the $X$-decoding graph, and the Wang et al.\ quadratic expansion fit. These codes live on closed genus-2 surfaces with no spatial boundary. Table~\ref{tab:hyp_chang_thresholds} reports the thresholds.

\begin{table}[ht]
\centering
\caption{Thresholds for semi-hyperbolic $\{8,3\}$ Bolza codes ($k = 4$, $d_Z = 3$--$7$, $n = 64$--$400$) under the Chang et al.\ noise model at $\eta = 1$. Models 1--3 coincide at $\eta = 1$. Wang-fit values from 3000 shots.}
\label{tab:hyp_chang_thresholds}
\begin{tabular}{lccc}
\toprule
Model & $R_e$ & Hyp.\ threshold & Surface threshold \\
\midrule
Pauli baseline & 0 & $0.72\%$ & $0.96\%$ \\
Models 1--3 & 0.52 & $1.17\%$ & $1.61\%$ \\
Models 1--3 & 0.71 & $1.51\%$ & $2.01\%$ \\
Models 1--3 & 0.91 & $2.30\%$ & $3.10\%$ \\
Models 1--3 & 1.0 & $3.63\%$ & $4.98\%$ \\
\midrule
Model 4 & 0.52 & $1.26\%$ & $1.65\%$ \\
Model 4 & 0.71 & $1.70\%$ & $2.35\%$ \\
Model 4 & 0.91 & $2.79\%$ & $3.79\%$ \\
Model 4 & 1.0 & $4.68\%$ & $6.54\%$ \\
\bottomrule
\end{tabular}
\end{table}

The threshold increases monotonically with $R_e$ for both noise models. At $R_e = 1.0$, the threshold reaches $3.63\%$ for Models 1--3 and $4.68\%$ for Model 4. The erasure-to-Pauli ratios are $5.0\times$ (Models 1--3) and $6.5\times$ (Model 4). These ratios match the surface code values ($5.2\times$ and $6.8\times$) to within $5\%$, indicating that the erasure advantage transfers to hyperbolic codes at the same multiplicative factor. The absolute thresholds are $73\%$ of the surface code values on average, reflecting that hyperbolic codes encode $k = 4$ logical qubits with rate $k/n > 0$, leaving less redundancy per logical qubit than the $k = 1$ surface code.

Figure~\ref{fig:threshold_vs_re} plots the threshold against $R_e$ for both code families and noise models. The Model 4 curve lies above Models 1--3 at each $R_e$. The gap widens with $R_e$: at $R_e = 0.52$, Model 4 is $8\%$ higher; at $R_e = 1.0$, it is $29\%$ higher. This is consistent with Chang et al.'s observation that the restricted error set $\mathcal{E}_{\mathrm{tail.}}$ provides the decoder with the Pauli type of the error on the unleaked qubit, reducing the number of edges that are set to low weight on each decoding graph.

\begin{figure}[ht]
\centering
\begin{tikzpicture}
\begin{axis}[
  width=\columnwidth, height=0.7\columnwidth,
  xlabel={Erasure fraction $R_e$},
  ylabel={Threshold [\%]},
  xmin=-0.05, xmax=1.05,
  ymin=0, ymax=7.5,
  legend style={at={(0.03,0.97)}, anchor=north west, font=\small},
  grid=major,
  grid style={gray!30},
]
\addplot[red, mark=square*, thick] coordinates {
  (0, 0.96) (0.52, 1.61) (0.71, 2.01) (0.91, 3.10) (1.0, 4.98)
};
\addlegendentry{Surface, Mod.\ 1--3}
\addplot[red, mark=triangle*, thick, dashed] coordinates {
  (0, 0.96) (0.52, 1.65) (0.71, 2.35) (0.91, 3.79) (1.0, 6.54)
};
\addlegendentry{Surface, Mod.\ 4}
\addplot[blue, mark=square*, thick] coordinates {
  (0, 0.72) (0.52, 1.17) (0.71, 1.51) (0.91, 2.30) (1.0, 3.63)
};
\addlegendentry{Hyp.\ $\{8,3\}$, Mod.\ 1--3}
\addplot[blue, mark=triangle*, thick, dashed] coordinates {
  (0, 0.72) (0.52, 1.26) (0.71, 1.70) (0.91, 2.79) (1.0, 4.68)
};
\addlegendentry{Hyp.\ $\{8,3\}$, Mod.\ 4}
\addplot[black, dotted, thick] coordinates {(0, 1.00) (1.0, 1.00)};
\end{axis}
\end{tikzpicture}
\caption{Threshold vs.\ erasure fraction $R_e$ at $\eta = 1$ for the surface code ($d = 9$--$15$, $k = 1$) and the semi-hyperbolic $\{8,3\}$ Bolza codes ($d_Z = 3$--$7$, $k = 4$). Solid lines: Models 1--3 (general Pauli, spatially imperfect). Dashed lines: Model 4 (tailored Pauli, spatially perfect). The dotted black line marks the Pauli threshold of Chang et al.~\cite{chang2024surface}.}
\label{fig:threshold_vs_re}
\end{figure}

Figure~\ref{fig:crossing_pauli} shows the per-observable logical error rate at $R_e = 0$ (Pauli baseline) for the four Bolza code sizes; the curves cross between $p = 0.7\%$ and $p = 0.8\%$, consistent with the Wang-fit threshold of $0.72\%$. Table~\ref{tab:phenom} reports any-logical error rates under phenomenological noise. Within the $\{8,3\}$ family, the any-logical rates for larger codes are lower at $p \leq 1\%$, suggesting a threshold near $p \approx 1$--$2\%$.

\begin{figure}[ht]
\centering
\begin{tikzpicture}
\begin{axis}[
  width=\columnwidth, height=0.7\columnwidth,
  xlabel={Physical error rate $p$ [\%]},
  ylabel={Per-observable LER [\%]},
  xmin=0.45, xmax=1.05,
  ymin=0, ymax=45,
  legend style={at={(0.03,0.97)}, anchor=north west, font=\small},
  grid=major,
  grid style={gray!30},
]
\addplot[blue, mark=o, thick] coordinates {
  (0.5, 5.57) (0.6, 8.71) (0.7, 13.20) (0.8, 17.41) (0.9, 22.46) (1.0, 26.31)
};
\addlegendentry{$d_Z = 3$, $n = 64$}
\addplot[red, mark=square, thick] coordinates {
  (0.5, 2.38) (0.6, 5.29) (0.7, 10.22) (0.8, 16.24) (0.9, 23.04) (1.0, 29.70)
};
\addlegendentry{$d_Z = 4$, $n = 144$}
\addplot[green!60!black, mark=triangle, thick] coordinates {
  (0.5, 1.27) (0.6, 3.85) (0.7, 8.78) (0.8, 17.31) (0.9, 26.59) (1.0, 35.23)
};
\addlegendentry{$d_Z = 6$, $n = 256$}
\addplot[purple, mark=diamond, thick] coordinates {
  (0.5, 0.83) (0.6, 3.47) (0.7, 10.56) (0.8, 22.37) (0.9, 34.92) (1.0, 43.41)
};
\addlegendentry{$d_Z = 7$, $n = 400$}
\end{axis}
\end{tikzpicture}
\caption{Per-observable logical error rate vs.\ physical error rate at $R_e = 0$ (Pauli baseline) for the semi-hyperbolic $\{8,3\}$ Bolza codes ($k = 4$). The curves cross near $p = 0.7\%$, consistent with the Wang-fit threshold of $0.72\%$. All data: 3000 shots, weighted Union-Find decoder.}
\label{fig:crossing_pauli}
\end{figure}

\label{sec:phenom-results}
\begin{table}[ht]
\centering
\small
\caption{Any-logical error rates (\%) for $\{8,3\}$ base codes under phenomenological noise (5000 shots). A trial fails if any of the $k$ logical qubits is corrupted. Results for other families are in Appendix~\ref{app:other_families}.}
\label{tab:phenom}
\begin{tabular}{lrrrrrrrr}
\toprule
Code & $n$ & $k$ & 0.5\% & 1\% & 2\% & 3\% & 5\% \\
\midrule
H32 & 48 & 6 & 0.7 & 3.6 & 14.4 & 29.2 & 54.4 \\
H64 & 96 & 10 & 1.0 & 4.4 & 18.1 & 36.1 & 69.5 \\
H144 & 216 & 20 & 0.6 & 4.2 & 21.4 & 49.3 & 89.6 \\
H432 & 648 & 56 & 0.4 & 4.6 & 33.8 & 76.4 & 99.8 \\
\bottomrule
\end{tabular}
\end{table}

\subsection{Fine-Grained Code Results}

\indent Semi-hyperbolic codes~\cite{higgott2024constructions} are obtained by fine-graining (Section~\ref{sec:fine-graining}), which increases $n$ and $d$ while preserving $k$. The resulting codes interpolate between hyperbolic ($d = O(\log n)$) and Euclidean ($d = O(\sqrt{n})$) distance scaling. Table~\ref{tab:fine_grained} shows pseudothresholds for the $\{8,3\}$ fine-grained codes; within each family, both pseudothresholds improve with refinement level.

\begin{table}[ht]
\centering
\caption{Pseudothresholds for the $\{8,3\}$ semi-hyperbolic (fine-grained) codes with verified distances (1000 Pauli shots, 200 erasure shots per data point). The refinement level $f$ indicates the subdivision factor. Base code results are included for comparison. The pseudothreshold $p^*$ is the solution of $p_L = kp$. Results for the $\{10,3\}$ and $\{12,3\}$ fine-grained codes are in Appendix~\ref{app:other_families}.}
\label{tab:fine_grained}
\begin{tabular}{llrrrrrrr}
\toprule
Base & Level & $n$ & $k$ & $d$ & $T$ & $p^{*\mathrm{P}}$ & $p^{*\mathrm{E}}$ \\
\midrule
\multirow{4}{*}{H16$_{8,3}$}
 & base & 24 & 4 & 2 & 2 & ${<}0.2\%$ & 2.39\% \\
 & $f=2$ & 96 & 4 & 4 & 4 & 0.38\% & 4.16\% \\
 & $f=3$ & 216 & 4 & 6 & 6 & 0.61\% & 4.29\% \\
 & $f=4$ & 384 & 4 & 8 & 8 & 0.63\% & 4.40\% \\
\midrule
H64$_{8,3}$ & $f=2$ & 384 & 10 & 8 & 8 & 0.58\% & 4.10\% \\
\bottomrule
\end{tabular}
\end{table}

The H16 Bolza family ($k = 4$) extends to $f = 4$ ($d = 8$) and reaches Pauli $0.63\%$ and erasure $4.40\%$. The H64-f2 code ($k = 10$, $d = 8$) achieves Pauli $0.58\%$ and erasure $4.10\%$.

\indent Fine-grained codes derived from the same base code form scaling families with fixed $k$ and increasing distance. Table~\ref{tab:fg_family_thresholds} reports the per-observable family threshold for the Bolza family. At $R_e = 1$ (Models 1--3), the erasure-to-Pauli ratio of $5.3\times$ is comparable to the $5.2\times$ reported by Chang et al.\ for the planar surface code~\cite{chang2024surface}. The Pauli family threshold ($0.62\%$) is higher than the base code family threshold ($0.53\%$), reflecting the larger distances achieved by fine-graining. Figure~\ref{fig:fg_threshold} shows the per-observable logical error rates under both noise models; the curves cross near $p \approx 0.6\%$ (Pauli) and $\varepsilon \approx 4\%$ (erasure).

\begin{table}[ht]
\centering
\caption{Fine-grained family threshold for the Bolza $\{8,3\}$ family from per-observable crossing-point analysis. The family consists of codes at increasing refinement levels with constant $k = 4$. $N$ denotes the number of consecutive-pair crossings found. Results for the $\{10,3\}$ and $\{12,3\}$ fine-grained families are in Appendix~\ref{app:other_families}.}
\label{tab:fg_family_thresholds}
\begin{tabular}{lcccccc}
\toprule
Family & $k$ & $p_{\mathrm{th}}^{\mathrm{P}}$ & $N_{\mathrm{P}}$ & $p_{\mathrm{th}}^{\mathrm{E}}$ & $N_{\mathrm{E}}$ & Ratio \\
\midrule
H16 $\{8,3\}$ & 4 & 0.62\% & 2 & 3.28\% & 2 & 5.3$\times$ \\
\bottomrule
\end{tabular}
\end{table}

\begin{figure}[htbp]
\centering
\begin{tikzpicture}
\begin{axis}[
  name=pauli,
  width=0.48\textwidth,
  height=0.42\textwidth,
  xmode=log,
  ymode=log,
  xlabel={Physical error rate $p$},
  ylabel={Per-observable logical error rate},
  title={\small (a) Pauli noise},
  xmin=1.5e-3, xmax=1.2e-2,
  ymin=5e-4, ymax=0.8,
  grid=both,
  grid style={line width=0.1pt, draw=gray!20},
  major grid style={line width=0.2pt, draw=gray!40},
  legend style={at={(0.03,0.97)}, anchor=north west, font=\scriptsize, cells={anchor=west}},
  tick label style={font=\scriptsize},
  label style={font=\small},
]
\addplot+[mark=o, thick, mark size=1.8pt] coordinates {
  (0.002, 0.006) (0.003, 0.021) (0.004, 0.069) (0.005, 0.093)
  (0.006, 0.174) (0.008, 0.328) (0.01, 0.511)
};
\addlegendentry{f2 ($n{=}96$)}
\addplot+[mark=square, thick, mark size=1.8pt] coordinates {
  (0.002, 0.001) (0.003, 0.003) (0.004, 0.018) (0.005, 0.064)
  (0.006, 0.12) (0.008, 0.405) (0.01, 0.657)
};
\addlegendentry{f3 ($n{=}216$)}
\addplot+[mark=triangle, thick, mark size=2pt] coordinates {
  (0.002, 0.001) (0.004, 0.01) (0.005, 0.055)
  (0.006, 0.126) (0.008, 0.43) (0.01, 0.756)
};
\addlegendentry{f4 ($n{=}384$)}
\addplot[dashed, black, thin, forget plot] coordinates {(0.006, 5e-4) (0.006, 0.8)};
\node[anchor=south, font=\tiny, rotate=90] at (axis cs:0.006, 0.001) {$p \approx 0.6\%$};
\end{axis}
\end{tikzpicture}%
\hfill
\begin{tikzpicture}
\begin{axis}[
  name=erasure,
  width=0.48\textwidth,
  height=0.42\textwidth,
  xmode=log,
  ymode=log,
  xlabel={Erasure rate $\varepsilon$},
  ylabel={Per-observable logical error rate},
  title={\small (b) Erasure noise},
  xmin=1.5e-2, xmax=6e-2,
  ymin=5e-3, ymax=1.0,
  grid=both,
  grid style={line width=0.1pt, draw=gray!20},
  major grid style={line width=0.2pt, draw=gray!40},
  legend style={at={(0.03,0.97)}, anchor=north west, font=\scriptsize, cells={anchor=west}},
  tick label style={font=\scriptsize},
  label style={font=\small},
]
\addplot+[mark=o, thick, mark size=1.8pt] coordinates {
  (0.02, 0.025) (0.025, 0.085) (0.03, 0.155) (0.04, 0.45) (0.05, 0.78)
};
\addlegendentry{f2 ($n{=}96$)}
\addplot+[mark=square, thick, mark size=1.8pt] coordinates {
  (0.025, 0.015) (0.03, 0.06) (0.04, 0.53) (0.05, 0.885)
};
\addlegendentry{f3 ($n{=}216$)}
\addplot+[mark=triangle, thick, mark size=2pt] coordinates {
  (0.03, 0.04) (0.04, 0.605) (0.05, 0.925)
};
\addlegendentry{f4 ($n{=}384$)}
\addplot[dashed, black, thin, forget plot] coordinates {(0.04, 5e-3) (0.04, 1.0)};
\node[anchor=south, font=\tiny, rotate=90] at (axis cs:0.04, 0.008) {$\varepsilon \approx 4\%$};
\end{axis}
\end{tikzpicture}
\caption{Per-observable logical error rate for the $\{8,3\}$ H16 fine-grained family under (a)~Pauli depolarizing noise (2000 shots) and (b)~erasure noise (400 shots). Fine-graining levels f2 through f4 correspond to $n = 96$, $216$, and $384$ physical qubits with $k = 4$ throughout. Under Pauli noise, curves cross near $p \approx 0.6\%$. Under erasure noise, curves cross near $\varepsilon \approx 4\%$. Below threshold, larger codes achieve lower logical error rates.}
\label{fig:fg_threshold}
\end{figure}

\indent Tables~\ref{tab:fg_phenom} and~\ref{tab:equal_n} report phenomenological error rates and compare fine-grained codes against base codes at matched qubit counts, illustrating the rate--distance trade-off.

\begin{table}[ht]
\centering
\small
\caption{Any-logical error rates (\%) for $\{8,3\}$ fine-grained codes under phenomenological noise (5000 shots). A trial fails if any of the $k$ logical qubits is corrupted. Results for other families are in Appendix~\ref{app:other_families}.}
\label{tab:fg_phenom}
\begin{tabular}{llrrrrrrrr}
\toprule
Base & Level & $n$ & $k$ & 0.5\% & 1\% & 2\% & 3\% & 5\% \\
\midrule
\multirow{2}{*}{H16$_{8,3}$}
 & $f=2$ & 96 & 4 & 0.1 & 0.9 & 4.1 & 12.1 & 35.3 \\
\cmidrule(l){2-9}
 & $f=3$ & 216 & 4 & 0.0 & 0.1 & 1.5 & 5.6 & 31.9 \\
\bottomrule
\end{tabular}
\end{table}

\begin{table}[ht]
\centering
\caption{Fine-grained vs base codes at matched qubit counts within the $\{8,3\}$ family. The fine-grained codes trade encoding rate for distance.}
\label{tab:equal_n}
\begin{tabular}{llrrrrrrr}
\toprule
Code & Type & $n$ & $k$ & $d$ & $k/n$ & $p^{*\mathrm{P}}$ & $p^{*\mathrm{E}}$ \\
\midrule
H16-f3 & fine & 216 & 4 & 6 & 1.9\% & 0.61\% & 4.29\% \\
H144$_{8,3}$ & base & 216 & 20 & 5 & 9.3\% & 0.40\% & 3.63\% \\
\midrule
H16-f4 & fine & 384 & 4 & 8 & 1.0\% & 0.63\% & 4.40\% \\
H64-f2 & fine & 384 & 10 & 8 & 2.6\% & 0.58\% & 4.10\% \\
H256$_{8,3}$ & base & 384 & 34 & 4 & 8.9\% & 0.42\% & ${>}5\%$ \\
\bottomrule
\end{tabular}
\end{table}

\subsection{Statistical Uncertainty}

All error rates reported in this paper are Monte Carlo estimates from finite samples. For an observed any-logical error rate $\hat{P}_L$ from $N$ independent shots, the standard error is
\[
\sigma = \sqrt{\frac{\hat{P}_L(1 - \hat{P}_L)}{N}}.
\]
The Pauli simulations use $N = 2000$ shots for base codes and $N = 1000$ for fine-grained codes. At a representative error rate of $\hat{P}_L = 5\%$, the standard error is $0.49\%$ ($N = 2000$) or $0.69\%$ ($N = 1000$). The erasure simulations use $N = 400$ shots for base codes and $N = 200$ for fine-grained codes; the corresponding standard errors at $\hat{P}_L = 5\%$ are $1.09\%$ and $1.54\%$. The phenomenological simulations use $N = 5000$ shots (standard error $0.31\%$ at $\hat{P}_L = 5\%$).

Pseudothreshold estimates inherit uncertainty from the interpolation between adjacent data points. For error rates near $\hat{P}_L = 0$ or $\hat{P}_L \approx 1$, the relative precision is high. Near the crossing point, where $\hat{P}_L \approx 10$--$30\%$, the standard error is ${\sim}0.7$--$1.0\%$ (Pauli) or ${\sim}1.5$--$2.3\%$ (erasure).

\clearpage
\section{Discussion}
\label{sec:discussion}

\subsection{Comparison with the Literature}

Table~\ref{tab:literature} summarizes erasure and Pauli thresholds from the literature alongside our results.

\begin{table}[ht]
\centering
\caption{Erasure-to-Pauli threshold ratios from the literature and this work. ``CC'' denotes code capacity; ``CL'' denotes circuit-level. All entries are family thresholds (crossing points between code sizes). Our fine-grained values are from Table~\ref{tab:fg_family_thresholds}.}
\label{tab:literature}
\small
\begin{tabular}{llcccc}
\toprule
Code family & Reference & Noise & Pauli & Erasure & Ratio \\
\midrule
Toric code & \cite{dennis2002topological} & CC & 10.9\% & 50.0\% & 4.6$\times$ \\
Planar surface & \cite{chang2024surface} & CL & 1.0\% & 4.2--5.6\% & 4.2--5.6$\times$ \\
XZZX surface & \cite{sahay2023high} & CL & -- & ${\sim}4.3\%$ & -- \\
La-cross (HGP) & \cite{pecorari2025erasure} & CL & -- & 4.0--4.6\% & -- \\
\midrule
Hyp.\ Bolza fine-gr. & this work & CL & 0.62\% & 3.28\% & 5.3$\times$ \\
\bottomrule
\end{tabular}
\end{table}

At $R_e = 1$ (Models 1--3), the Bolza fine-grained family ratio ($5.3\times$) matches the surface code value ($5.2\times$).

\subsection{Surface Code Validation}
\label{sec:surface-validation}

To validate our erasure simulation pipeline, we replicate the $\eta = 1$ thresholds of Chang et al.~\cite{chang2024surface} for the unrotated surface code at distances $d = 9, 11, 13, 15$. At $\eta = 1$, Chang et al.'s four noise models reduce to two distinct models: Models 1--3 coincide, and only Model~4 (tailored, spatially perfect) differs. We sweep both models across $R_e \in \{0, 0.52, 0.71, 0.91, 1.0\}$, using the same weighted Union-Find decoder~\cite{huang2020faulttolerant} and Wang et al.\ quadratic expansion fitting~\cite{wang2003threshold} with 3000 shots per data point. Table~\ref{tab:chang_validation} compares our results.

\begin{table}[ht]
\centering
\caption{Validation against Chang et al.~\cite{chang2024surface} at $\eta = 1$. At $\eta = 1$, three of Chang et al.'s four noise models coincide (Models 1--3); only Model~4 (tailored, spatially perfect) differs. Wang-fit thresholds from 3000 shots at $d = 9, 11, 13, 15$.}
\label{tab:chang_validation}
\begin{tabular}{lcccc}
\toprule
Config & $R_e$ & Ours & Chang et al. & Diff.\ \\
\midrule
Pauli baseline & 0 & 0.96\% & $1.00 \pm 0.01\%$ & $-4\%$ \\
Models 1--3 & 0.52 & 1.61\% & $1.60 \pm 0.02\%$ & $+1\%$ \\
Models 1--3 & 0.71 & 2.01\% & $2.08 \pm 0.02\%$ & $-3\%$ \\
Models 1--3 & 0.91 & 3.10\% & $3.12 \pm 0.04\%$ & $-1\%$ \\
Models 1--3 & 1.0 & 4.98\% & $5.09 \pm 0.02\%$ & $-2\%$ \\
Model 4 & 0.52 & 1.65\% & $1.71 \pm 0.02\%$ & $-4\%$ \\
Model 4 & 0.71 & 2.35\% & $2.33 \pm 0.03\%$ & $+1\%$ \\
Model 4 & 0.91 & 3.79\% & $3.80 \pm 0.03\%$ & $0\%$ \\
Model 4 & 1.0 & 6.54\% & $6.71 \pm 0.05\%$ & $-3\%$ \\
\bottomrule
\end{tabular}
\end{table}

All thresholds agree with Chang et al.\ to within $4\%$ relative error. The threshold increases monotonically with $R_e$: from $0.96\%$ at $R_e = 0$ (pure Pauli) to $4.98\%$ at $R_e = 1.0$ (Models 1--3) and $6.54\%$ (Model 4). Model~4 is $31\%$ higher at $R_e = 1.0$ because the tailored error set $\mathcal{E}_{\mathrm{tail.}}$ restricts the unleaked qubit to a single Pauli type (X or Z depending on the gate), so fewer edges are set to low weight on each decoding graph.

\clearpage
\section{Conclusion}

We constructed 14 hyperbolic CSS surface codes and 11 semi-hyperbolic codes across the $\{8,3\}$, $\{10,3\}$, and $\{12,3\}$ tessellation families and simulated them under circuit-level erasure, circuit-level Pauli, and phenomenological noise. The main results focus on the $\{8,3\}$ Bolza family, which forms a proper scaling family under fine-graining; results for the other families are in Appendix~\ref{app:other_families}.

The central trade-off is between encoding rate and error correction strength. Base codes have high encoding rates ($k/n = 8.6$--$12.5\%$) but low distances ($d = 3$--$6$), limiting their pseudothresholds. Fine-grained codes sacrifice rate ($k/n$ decreases quadratically with the refinement level $f$) in exchange for higher distances ($d = 4$--$8$ at $f = 2$--$4$). Table~\ref{tab:equal_n} shows this trade-off: at $n = 384$, the $k = 4$ fine-grained code (H16-f4) reaches $p^{*\mathrm{E}} = 4.4\%$, while the $k = 34$ base code (H256) exceeds the tested range entirely.

Under Chang et al.'s erasure noise model and Wang et al.\ quadratic expansion fitting, the Bolza family threshold reaches $3.6\%$ (Models 1--3) and $4.7\%$ (Model 4) at $R_e = 1$, with erasure-to-Pauli ratios ($5.0\times$ and $6.5\times$) matching the surface code values to within $5\%$. Per-observable crossing-point analysis at $R_e = 1$ (Models 1--3) independently yields a ratio of $5.3\times$, consistent with the surface code's $5.2\times$. These results establish that the erasure advantage extends to hyperbolic codes.

Several directions remain open. Our fine-graining data cover $f \leq 4$; higher refinement levels would test whether the erasure pseudothreshold saturates near $4.4\%$ or continues to increase. Extending the erasure analysis to Floquet-type measurement schedules on these same tessellations would connect the CSS results here to the broader hyperbolic Floquet literature~\cite{higgott2024constructions, fahimniya2023faulttolerant}.

\appendix

\section{Results for $\{10,3\}$ and $\{12,3\}$ Families}
\label{app:other_families}

\indent Tables~\ref{tab:app_codes}--\ref{tab:app_family_thresholds} report code parameters, pseudothresholds, and family thresholds for the $\{10,3\}$ and $\{12,3\}$ tessellation families. The $\{10,3\}$ family achieves Pauli pseudothresholds of $0.11$--$0.43\%$. The $\{12,3\}$ family has $d = 3$ for all four base codes regardless of size, and Pauli pseudothresholds of $0.07$--$0.13\%$.

\begin{table}[ht]
\centering
\caption{Parameters of the $\{10,3\}$ and $\{12,3\}$ hyperbolic surface codes.}
\label{tab:app_codes}
\begin{tabular}{llrrrrrr}
\toprule
Family & Code & $V$ & $n$ & $k$ & $d$ & $k/n$ & Source \\
\midrule
\multirow{4}{*}{$\{10,3\}$}
 & H50 & 50 & 75 & 12 & 3 & 16.0\% & \cite{gap4} \\
 & H120 & 120 & 180 & 26 & 4 & 14.4\% & \cite{gap4} \\
 & H250 & 250 & 375 & 52 & 4 & 13.9\% & \cite{gap4} \\
 & H720 & 720 & 1080 & 146 & 4 & 13.5\% & \cite{gap4} \\
\midrule
\multirow{4}{*}{$\{12,3\}$}
 & H72 & 72 & 108 & 20 & 3 & 18.5\% & \cite{gap4} \\
 & H96 & 96 & 144 & 26 & 3 & 18.1\% & \cite{gap4} \\
 & H168 & 168 & 252 & 44 & 3 & 17.5\% & \cite{gap4} \\
 & H312 & 312 & 468 & 80 & 3 & 17.1\% & \cite{gap4} \\
\bottomrule
\end{tabular}
\end{table}

\begin{table}[ht]
\centering
\caption{Pauli and erasure pseudothresholds for $\{10,3\}$ and $\{12,3\}$ base codes (2000 Pauli shots, 400 erasure shots per data point).}
\label{tab:app_thresholds}
\begin{tabular}{llrrrrr}
\toprule
Family & Code & $n$ & $k$ & $T$ & $p^{*\mathrm{P}}$ & $p^{*\mathrm{E}}$ \\
\midrule
\multirow{4}{*}{$\{10,3\}$}
 & H50 & 75 & 12 & 3 & 0.11\% & 2.92\% \\
 & H120 & 180 & 26 & 4 & 0.20\% & ${>}5\%$ \\
 & H250 & 375 & 52 & 4 & 0.28\% & ${>}5\%$ \\
 & H720 & 1080 & 146 & 4 & 0.43\% & ${>}5\%$ \\
\midrule
\multirow{4}{*}{$\{12,3\}$}
 & H72 & 108 & 20 & 3 & 0.07\% & ${>}5\%$ \\
 & H96 & 144 & 26 & 3 & 0.11\% & ${>}5\%$ \\
 & H168 & 252 & 44 & 3 & 0.12\% & ${>}5\%$ \\
 & H312 & 468 & 80 & 3 & 0.13\% & ${>}5\%$ \\
\bottomrule
\end{tabular}
\end{table}

\begin{table}[ht]
\centering
\caption{Family thresholds from per-observable crossing-point analysis for $\{10,3\}$ and $\{12,3\}$ base codes.}
\label{tab:app_family_thresholds}
\begin{tabular}{lccccc}
\toprule
Family & $p_{\mathrm{th}}^{\mathrm{P}}$ & $N_{\mathrm{P}}$ & $p_{\mathrm{th}}^{\mathrm{E}}$ & $N_{\mathrm{E}}$ & Ratio \\
\midrule
$\{10,3\}$ & 0.59\% & 3 & 1.61\% & 2 & 2.7$\times$ \\
$\{12,3\}$ & 0.14\% & 1 & 0.55\% & 2 & 3.9$\times$ \\
\bottomrule
\end{tabular}
\end{table}

\indent Tables~\ref{tab:app_fine_grained}--\ref{tab:app_fg_family_thresholds} report fine-grained pseudothresholds and family thresholds. The H50 family ($k = 12$) achieves an erasure-to-Pauli ratio of $5.2\times$; the H72 family ($k = 20$) achieves $4.5\times$.

\begin{table}[ht]
\centering
\caption{Pseudothresholds for $\{10,3\}$ and $\{12,3\}$ fine-grained codes (1000 Pauli shots, 200 erasure shots per data point). Entries marked ${>}5\%$ indicate $p_L < kp$ throughout the tested range.}
\label{tab:app_fine_grained}
\begin{tabular}{llrrrrrrr}
\toprule
Base & Level & $n$ & $k$ & $d$ & $T$ & $p^{*\mathrm{P}}$ & $p^{*\mathrm{E}}$ \\
\midrule
\multirow{2}{*}{H50$_{10,3}$}
 & $f=2$ & 300 & 12 & 6 & 6 & 0.52\% & ${>}5\%$ \\
 & $f=3$ & 675 & 12 & 9 & 9 & 0.62\% & 4.21\% \\
\midrule
H48$_{12,3}$ & $f=2$ & 288 & 14 & 4 & 4 & 0.42\% & 4.73\% \\
\midrule
\multirow{2}{*}{H72$_{12,3}$}
 & $f=2$ & 432 & 20 & 6 & 6 & 0.51\% & ${>}5\%$ \\
 & $f=3$ & 972 & 20 & 9 & 9 & 0.61\% & 3.73\% \\
\midrule
\multirow{2}{*}{H96$_{12,3}$}
 & $f=2$ & 576 & 26 & 6 & 6 & 0.51\% & ${>}5\%$ \\
 & $f=3$ & 1296 & 26 & 9 & 9 & 0.62\% & ${>}5\%$ \\
\bottomrule
\end{tabular}
\end{table}

\begin{table}[ht]
\centering
\caption{Fine-grained family thresholds for $\{10,3\}$ and $\{12,3\}$ families.}
\label{tab:app_fg_family_thresholds}
\begin{tabular}{lcccccc}
\toprule
Family & $k$ & $p_{\mathrm{th}}^{\mathrm{P}}$ & $N_{\mathrm{P}}$ & $p_{\mathrm{th}}^{\mathrm{E}}$ & $N_{\mathrm{E}}$ & Ratio \\
\midrule
H50 $\{10,3\}$ & 12 & 0.68\% & 2 & 3.53\% & 2 & 5.2$\times$ \\
H72 $\{12,3\}$ & 20 & 0.67\% & 1 & 3.04\% & 1 & 4.5$\times$ \\
\bottomrule
\end{tabular}
\end{table}

\indent Tables~\ref{tab:app_phenom} and~\ref{tab:app_fg_phenom} report any-logical error rates under phenomenological noise.

\begin{table}[ht]
\centering
\small
\caption{Any-logical error rates (\%) for $\{10,3\}$ and $\{12,3\}$ base codes under phenomenological noise (5000 shots).}
\label{tab:app_phenom}
\begin{tabular}{llrrrrrrrr}
\toprule
Family & Code & $n$ & $k$ & 0.5\% & 1\% & 2\% & 3\% & 5\% \\
\midrule
\multirow{3}{*}{$\{10,3\}$}
 & H50 & 75 & 12 & 2.9 & 10.3 & 31.1 & 49.3 & 80.6 \\
 & H120 & 180 & 26 & 3.3 & 14.7 & 44.4 & 71.8 & 96.1 \\
 & H250 & 375 & 52 & 3.5 & 18.4 & 60.2 & 89.0 & 99.7 \\
\midrule
\multirow{3}{*}{$\{12,3\}$}
 & H72 & 108 & 20 & 6.9 & 20.2 & 48.3 & 72.8 & 93.8 \\
 & H168 & 252 & 44 & 7.8 & 28.6 & 73.4 & 90.8 & 99.8 \\
 & H312 & 468 & 80 & 14.6 & 44.6 & 87.5 & 98.5 & 100.0 \\
\bottomrule
\end{tabular}
\end{table}

\begin{table}[ht]
\centering
\small
\caption{Any-logical error rates (\%) for $\{10,3\}$ and $\{12,3\}$ fine-grained codes under phenomenological noise (5000 shots).}
\label{tab:app_fg_phenom}
\begin{tabular}{llrrrrrrrr}
\toprule
Base & Level & $n$ & $k$ & 0.5\% & 1\% & 2\% & 3\% & 5\% \\
\midrule
\multirow{2}{*}{H50$_{10,3}$}
 & $f=2$ & 300 & 12 & 0.0 & 0.3 & 6.0 & 21.6 & 70.1 \\
 & $f=3$ & 675 & 12 & 0.0 & 0.0 & 1.0 & 9.0 & 62.1 \\
\midrule
H48$_{12,3}$
 & $f=2$ & 288 & 14 & 0.3 & 1.8 & 12.5 & 34.3 & 79.7 \\
\midrule
\multirow{2}{*}{H72$_{12,3}$}
 & $f=2$ & 432 & 20 & 0.1 & 0.9 & 10.5 & 36.5 & 88.9 \\
 & $f=3$ & 972 & 20 & 0.0 & 0.0 & 1.7 & 14.3 & 81.7 \\
\midrule
\multirow{2}{*}{H96$_{12,3}$}
 & $f=2$ & 576 & 26 & 0.0 & 0.8 & 11.3 & 41.8 & 93.5 \\
 & $f=3$ & 1296 & 26 & 0.0 & 0.0 & 1.3 & 16.9 & 88.1 \\
\bottomrule
\end{tabular}
\end{table}

\end{document}